# A Visual Analytics Based Decision Making Environment for COVID-19 Modeling and Visualization

Shehzad Afzal*
KAUST

Sohaib Ghani†
KAUST

Hank C. Jenkins-Smith‡
University of Oklahoma

David S. Ebert §
Purdue University

Markus Hadwiger ¶
KAUST

Ibrahim Hoteit∥
KAUST

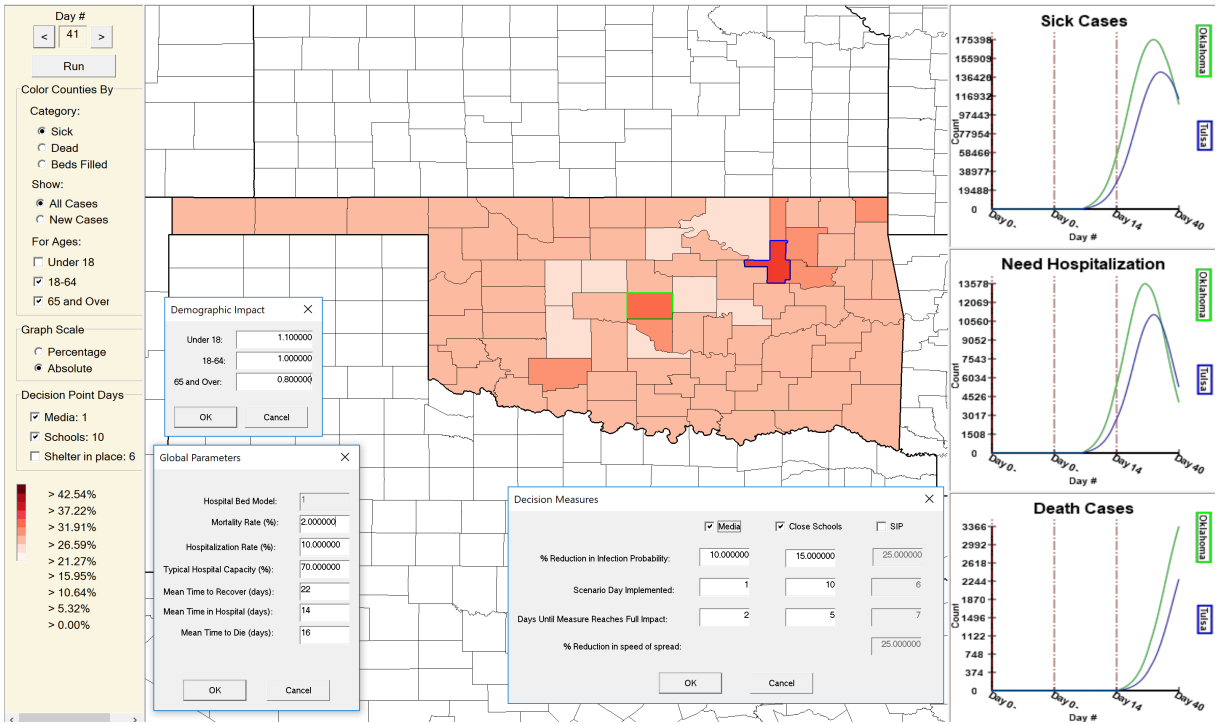

Figure 1: Visual analytics environment for COVID-19 modeling, visualization, and decision support. The map view shows the status of the spread on day 41 of simulation in the Oklahoma State. Detailed statistics (number of sick cases, people that need hospitalization, and deaths) for the selected counties (Oklahoma City and Tulsa) are shown on the right. Decision measures 'media alerts' and 'school closures' are applied on day 1 and day 10, respectively. Interface to configure model parameters is also shown.

## ABSTRACT

Public health officials dealing with pandemics like COVID-19 have to evaluate and prepare response plans. This planning phase requires not only looking into the spatiotemporal dynamics and impact of the pandemic using simulation models, but they also need to plan and ensure the availability of resources under different spread scenarios. To this end, we have developed a visual analytics environment that enables public health officials to model, simulate, and explore the spread of COVID-19 by supplying county-level information such as population, demographics, and hospital beds. This environment facilitates users to explore spatiotemporal model simulation data relevant to COVID-19 through a geospatial map with linked statistical views, apply different decision measures at different points in time, and understand their potential impact. Users can drill-down to county-level details such as the number of sicknesses, deaths, needs for hospitalization, and variations in these statistics over time. We demonstrate the usefulness of this environment through a use case study and also provide feedback from domain experts. We also provide details about future extensions and potential applications of this work.

**Index Terms:** Human-centered computing—Visualization—Visualization application domains—Visual analytics

## 1 INTRODUCTION

Public health officials often need to analyze simulations of epidemic models to improve preparedness, planning responses, and mitigate the impact of pandemics like COVID-19. During the planning and preparatory stages, they need to explore different scenarios and decision measures, and study the impact of these decision measures on controlling the pandemic's spread. They also have to analyze the availability of different resources during the various phases of an epidemic outbreak keeping in view the outputs from simulation models and their current stocks, and make plans accordingly. In their analysis tasks, they often need to explore simulation outputs in a spatiotemporal environment where they can analyze the spread patterns and the evolution of spread across space and time, interactively explore and filter, apply decision measures and analyze their impact at different points in space and time, modify model parameters, and explore different scenarios.

*e-mail: shehzad.afzal@kaust.edu.sa
†e-mail: sohaib.ghani@kaust.edu.sa
‡e-mail: hjsmith@ou.edu
§e-mail: ebertd@purdue.edu
¶e-mail: markus.hadwiger@kaust.edu.sa
∥e-mail: ibrahim.hoteit@kaust.edu.sa

The goal of this work is to provide a visual analytics environment to facilitate COVID-19 modeling, exploration, and visualization. The analysts and public health officials can provide county-level information such as population statistics, demographics, medical resources, and analyze the model simulation results in multiple linked views of our visual analytics environment. The environment enables spatiotemporal exploration of model simulation outputs, and supports users to apply decision measures at different points in time to create and study different spread scenarios and evaluate the response effectiveness in containing the spread. This work is an extension of our previous epidemic modeling work related to Pandemic Influenza and Rift Valley Fever [4, 19]. The main contributions of our work are as follows. 1) COVID-19 modeling and simulation, extending our work on Pandemic Influenza [4, 19], 2) A visual analytics environment where users can explore COVID-19 spread scenarios, apply different decision measures to create scenarios, adjust model parameters, and prepare plans to ensure availability of resources. This work is completed in collaboration with domain experts and scientists working in public health, epidemic modeling, emergency response, and decision support to facilitate COVID-19 response.

## 2 RELATED WORK

Several visual analytics and visualization based works have been presented in the last few years for analyzing epidemic or pandemic data [10, 15]. Preim and Lawonn [23] recently published a detailed survey of visual analytics tools designed for the management of infectious diseases, chronic diseases, and other healthcare issues.

One major area for the management of infectious disease is to design an interactive spread model with various parameters. Liang et al. [18] designed a visualization based malaria spread model to understand the spread of malaria under different climate conditions. Maciejewski et al. [19] also designed a tool kit for understanding the spread of influenza. Recently, various tools and models have been implemented to model the spread of COVID-19 under various conditions [11, 12, 14, 20, 26, 28, 30]. Predictive analytics is often used to understand and forecast epidemic spread [5, 7, 9, 16, 21]. Researchers have also utilized social media data to understand and predict the spread of epidemics [6, 24, 27]. Mobility data have also been used to study the outbreak of epidemics in certain areas [8, 32].

Visual analytics based decision support tools are generally designed to allow decision-makers to interactively change various conditions and parameters and visualize outputs for making well-informed decisions during epidemics or pandemics. PandemCap [29] is a decision support tool designed to help public health decision-makers in studying the impact of various control measures such as hospital beds required. Afzal et al. [4] introduced a decision history visualization and navigation tool that enables analysis, management, and comparison between different epidemic spread scenarios and decision measures. Recently, various decision support tools have been designed for better management and mitigating the outcomes of the COVID-19 [13, 22, 25, 31]. These tools are designed for analyzing specific parameters and geolocations and lack detailed visual analytics support for decision making. In this work, we present an interactive visual analytics-based decision support environment designed in collaboration with domain experts to help make better decisions to manage COVID-19 situation.

## 3 VISUAL ANALYTICS ENVIRONMENT

Our visual analytics environment consists of multiple linked views comprising the geospatial map view, time series visualizations, decision measures and filtering options, and time scroller to enable users to display simulation data at different points in time. The visual analytics environment is shown in Fig. 1. Users can interactively explore the simulation data produced by the COVID-19 model in the map view. Users can scroll through time and analyze the COVID-19 spread through the color-coded geographical representation of the simulation model output. Users can select different counties on the map, and corresponding time series are loaded in the linked time series visualizations showing the number of sick cases, deaths, and the number of people that may need hospitalization (Fig.1(right)). This enables users to compare the statistics across multiple regions on the map.

Users can apply any decision measure (shelter in place, school closures, media campaigns) at any point in time by scrolling to that day value and then enabling the decision measure. The model then regenerates the simulation outputs incorporating that decision measure. Parameters relevant to decision measures can be modified, such as the number of days required to reach the full impact of the decision measure, and decrease in the baseline prevalence due to these decision measures. Users can also make adjustments in the parameter settings of the model, location of the initial cases, enabling/disabling air transportation, spread rate, impact on different demographics, and other similar settings through the interface options shown in Fig. 1. The input to this visual analytics environment consists of county-level data that includes population, demographics, airport locations, and the number of hospital beds.

## 4 COVID-19 MODELING

We extended our prior work on Pandemic Influenza modeling [4, 19] to model COVID-19. Similar to our Pandemic Influenza model, it is a person-to-person contact model that is derived from compartment models traditionally used for modeling infectious diseases. In SIR compartment models [17], the entire population is divided into three compartments: Susceptible (S), Infectious (I), and Recovered (R) and the population transitions between these three compartments based on equations for disease dynamics. The COVID-19 model generates new sicknesses, deaths, and hospitalizations on a daily basis for each county. Although the environment itself is flexible to support any granularity level for spatial regions, but in this particular implementation county is used as the smallest spatial unit. The baseline prevalence curve that defines a probability of each person getting infected with COVID-19 is approximated based on basic reproduction number (Ro). This curve is then used to derive the model and generate daily infections for each county.

There are other parameters defined for the COVID-19 model. These parameters include mortality rate, time to death, recovery rate and time, hospitalization rate, number of days in the hospital, number of used beds, the incubation period (time from exposure to an infected person to show symptoms), and shedding period (time period during which an infected person remains contagious). There is a spread rate parameter that controls the spatial dynamics of COVID-19, controlling the speed at which it travels between different counties. Air travel can also be enabled or disabled during model simulation. We defined these parameters based on various studies and reported information [2, 3]. Users can create different scenarios by modifying different parameters of the model.

Since the impact of COVID-19 could be different depending on the age groups, the input population is divided into different age groups, and different baseline prevalence curve is defined for each group. Person-to-person contact rate and spread rate can also vary based on population density. In order to capture the variations in the disease spread dynamics due to population density, each county is classified into rural, small, or urban categories and a different disease spread rate modifier is then used for each population density class. Most of these model parameters can be modified through an interface shown in Fig. 1. Other parameters can be modified through input configuration files. Due to differences in population density among counties (corresponding spread rate modifier) and demographic variations (corresponding baseline prevalence curve), COVID-19 can peak differently in different counties even if the initial cases are defined at the same time. The number of available beds in hospitals also has an impact on the number of deaths, and this data defined at the county level is also passed as input to the model. By default, the hospitals are considered to be 70 percent occupied [19], but can be modified by the user according to the hospital beds utilization status in counties.

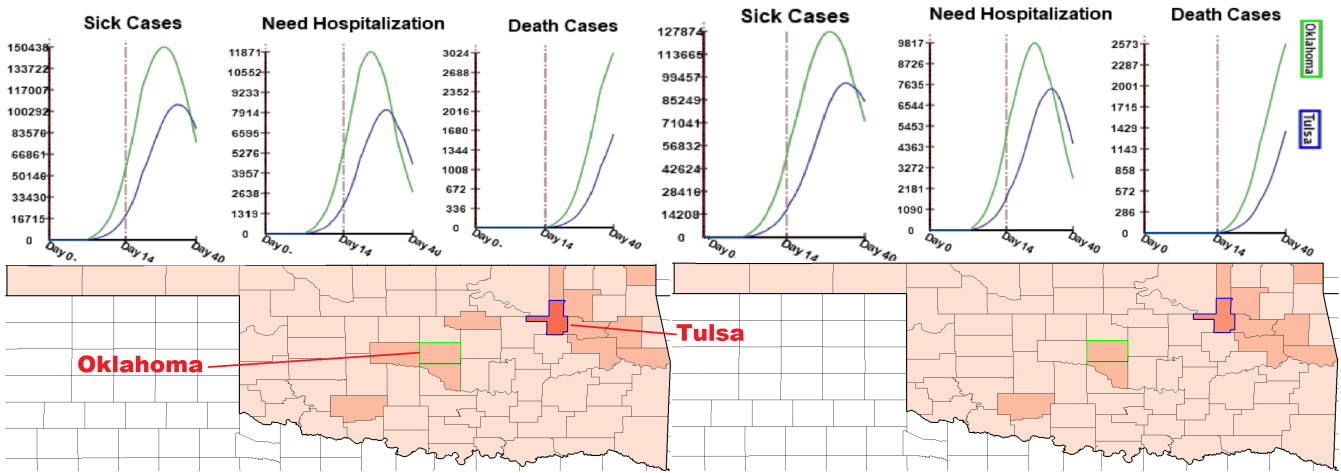

Figure 2: COVID-19 spread scenarios in Oklahoma State on day 41. (Left) Decision measure 'shelter in place' is applied on day 15, 'media alerts' on day 1, and 'school closures' on day 10. (Right) 'Shelter in place' is applied on day 10, 'media alerts' on day 1, and 'school closures' on day 5.

There are different decision measures that can be incorporated into the model calculations. Whenever a decision measure is applied at different temporal points in the model simulation, the baseline prevalence is reduced by a certain value that is predefined for each decision measure and can be adjusted by the user. The start time of the application of the decision measure, along with the number of days until the decision measure reaches the full impact, is also incorporated in the model calculations. The decision measures currently implemented in the system include school closures, media alerts, and shelter in place orders. Users can also create and explore different scenarios by using different combinations of these decision measures that can be applied at different points in time.

## 5 Use Case Scenarios

We discuss some use case scenarios to showcase how this visual analytics environment can be utilized by public health officials to prepare and plan their responses to COVID-19. Fig. 1 shows the COVID-19 pandemic spread in the Oklahoma State based on the simulation by the COVID-19 model originating from Oklahoma City. This image shows the total number of infections on day 41 of the simulation in the combined age groups of 18-64, and 65 and over. In this particular scenario, the analyst is interested in analyzing the total number of cases and how they change over time in Oklahoma City and Tulsa. There are three decision measures available for selection: media alerts, school closures, and shelter in place. During the simulation, the analyst has enabled the decision measure media alerts on day 1, and school closures on day 10 and kept shelter in place disabled. The map view shows the population percentage infected from COVID-19 and linked time series shows the number of sick, hospitalized, and death cases in the two selected cities. Air travel is also enabled in the spread simulation; that is why the infections appear very quickly in Tulsa that is directly connected to Oklahoma City through flights.

The analyst also wants to know the estimated duration and when the cases will peak, based on the given model parameters and settings, and selected decision measures. In this particular scenario, cases peak around day 30, and the duration of the spread is close to 75 days. The analyst then explores alternative scenarios such as the impact of enabling shelter in place decision measure on day 15. Fig. 2(left) shows that the number of sick cases are dropped by almost 25 thousand in Oklahoma City on day 41 when the cases hit the peak. Similarly, the analyst creates another scenario by selecting the same age group (18-64, and 65 and over) but excluding airports in the model simulation. Decision measures are also applied earlier in time like the previous case. Media alerts on day 1, school closures on day 5, and shelter in place on day 10. Fig. 2(right) shows the results of this simulation, and the number of infections near the peak are reduced by almost 50 thousand in Oklahoma City.

In the next scenario, the analyst is interested in the utilization of available beds during the course of the pandemic. Analyst selects the 'beds filled' category to be displayed on the map. All age groups are included while conducting this analysis. Decision measures media alerts and school closures are applied at day 1 and 5, respectively, while decision measure shelter in place is not applied. The analyst is particularly interested in examining the county resources (number of hospital beds) of Oklahoma City and Tulsa. Fig. 3 shows the status of the entire Oklahoma State on day 5 and day 9. Hospital capacity is set at 70% by default when the simulation is run. This is also configurable as the user can change the capacity to any other percentage value and then run the simulation. It can be seen in Fig. 3 that on day 5 Oklahoma City and many other counties reach their full capacity while Tulsa reaches around 80% of its total bed capacity. On day 9, Tulsa also needs additional beds, as almost 96% of these resources are utilized. This analysis helps the identification of stress on county medical resources and provides some insights about additional hospital beds needed to deal with this pandemic scenario.

## 6 Discussion and Planned Extensions

This work was initiated as an effort to help public health officials prepare and evaluate plans to deal with the COVID-19 pandemic. Public health officials often need to evaluate the effectiveness of different mitigative strategies before making policy decisions. One of the main limitations is the lack of interactive visual interfaces and visualization capabilities with existing models that can help them

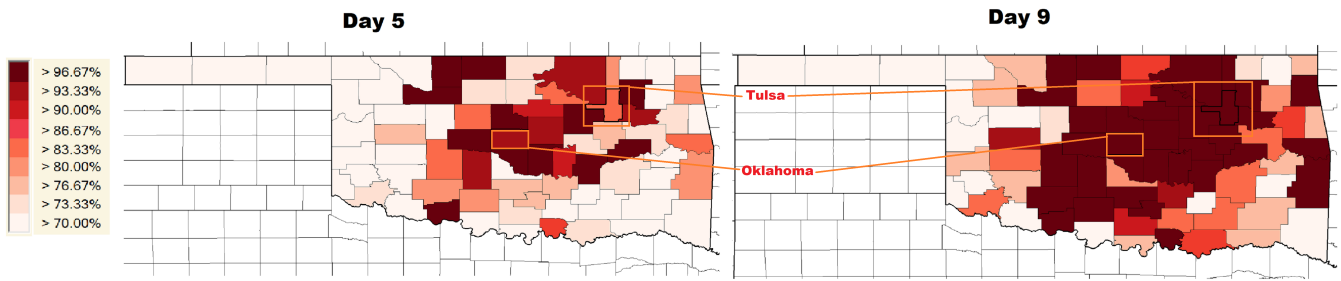

Figure 3: The simulation of the number of beds filled on day 5 (left) and day 9 (right) due to COVID-19. Decision measure 'media alerts' is applied on day 1, whereas 'school closures' is applied on day 5. Oklahoma City and Tulsa are selected on both maps. (Left) On day 5, Tulsa is at almost 80 percent capacity, whereas Oklahoma City is at almost 93 percent capacity. (Right) On day 9, both counties have reached full bed capacity.

explore different strategies and evaluate the responses.

We looked into public health officials' requirements who had similar constraints while dealing with the COVID-19 pandemic outbreak. One of their initial requirements was to get estimates about the number of sicknesses, number of hospitalizations, and duration of the pandemic. They also required a mechanism where they can interactively modify model parameters and settings to create and study different scenarios. In order to address these requirements, we extended our earlier work on pandemic influenza modeling [4, 19] to implement a visual analytics environment for modeling and visualization of COVID-19. Another requirement was to provide an additional decision measure 'shelter in place' where they can explore different scenarios by applying this measure at different points in time and evaluate the dynamics of the spread and their responses.

They were also interested in functionality to evaluate the availability of hospital beds throughout the pandemic wave. They were looking to input the total number of hospital beds at the county level and their availability status, and then use COVID-19 model simulations to estimate hospital beds' requirements in different outbreak scenarios. We are currently working on extensions to this functionality to support analyzing the availability of additional medical facilities supplies such as ICU's and ventilators at the county level. In these extensions, analysts would be able to enter the current number of these supplies at the county level and then analyze the utilization status at different times in simulation based on the baseline prevalence curve derived from the model.

Another desired feature was the capability to adjust the baseline prevalence curve that is used to derive the model and calculate the daily infections, which has a direct relation with the basic reproduction number (Ro). Once the user applies any decision measure, it adjusts the baselines' prevalence curve. This option is also configurable, enabling users to adjust each decision measure's impact in controlling the COVID-19 spread as shown in Fig. 1. This functionality also enables introducing additional decision measures, where adjustments in the baseline curve based on that decision measure can be driven by actual data or estimates.

The spatial dynamics of the disease are also controlled through a spread rate parameter that is configurable in the visual analytics environment. One of the planned extensions is to incorporate additional datasets like connectivity information (road networks and google maps traffic datasets) to dynamically adjust this attack rate parameter. We also plan to utilize Cuebiq mobility insights [1], such as Mobility Flows Analysis (quantifiable information about how the populations cross the county boundaries over time) and Cuebiq Mobility Index (CMI) (mobility trends patterns over time). These datasets can be utilized for data-driven adjustment in the baseline prevalence curve whenever a decision measure like 'shelter in place' is applied. Similarly, spread rate parameters can also be adjusted based on these datasets. These datasets are also useful to introduce additional decision measures in the model like social distancing, and its effectiveness can be derived from these datasets.

If an analyst wants to explore and compare multiple scenarios and combinations of decision measures at different points in time, there is a need to provide an interactive decision history tree [4] like data structure that facilitates management and comparison of different decision paths. We plan to extend our previous work [4] to adapt decision history visualization and navigation tool for COVID-19 modeling and visualization. We also plan to provide interactive selection and quarantine of spatial regions on the map and study the quarantine impact. We are also looking into datasets such as Cuebiq contact, visit, and mobility indexes [1] to incorporate additional decision measures, e.g., 'economic reopening.' To facilitate the task requirements to compare the impact of different decision measures applied on certain combination of spatial regions at different points in time, we will extend the decision history visualization and navigation tool [4] with a support for thumbnails-based visual comparison linked with nodes that represent decision measures.

This particular work is focused on analyzing the COVID-19 outbreak in Oklahoma State because public health officials were interested in using visual analytics-based COVID-19 tools that can facilitate understanding the spread dynamics, prepare plans, and explore different scenarios. The feedback was generally positive, and they found it useful in their analysis. They requested additional features, such as the ability to analyze the utilization of medical resources based on simulations in different spread scenarios. They were also interested in additional data import features and the ability to generate a combined summary for the entire state. We also plan to explore how model uncertainties can be captured and communicated to the decision makers through this visual analytics environment. The environment itself is scalable and can be applied to a much larger scale [19] provided that the required data is available.

## 7 CONCLUSION

This work presented a visual analytics-based environment for COVID-19 modeling, visualization, and decision support to help public health officials prepare and exercise response plans in pandemic outbreak scenarios. Utilizing this visual analytics environment, public health officials can explore COVID-19 county-level simulation data generated by our model, apply different decision measures to reduce the pandemic's impact, and modify model parameters and settings to create and analyze different spread scenarios. They can also use this environment to analyze the availability of resources like hospital beds during different phases of COVID-19 spread at the county level. We provided several use-case scenarios to demonstrate the effectiveness of the system.